\documentclass{aa}
\usepackage{psfig,latexsym,longtable,lscape}

\def\lsim{\lower0.5ex\hbox{$\; \buildrel < \over \sim \;$}}
\def\ros{{\sl ROSAT }}

\def\ein{{\sl Einstein }}

\def\it{\sl}
\def\asec{\ifmmode ^{\prime\prime}\else$^{\prime\prime}$\fi}
\def\amin{\ifmmode ^{\prime}\else$^{\prime}$\fi}

\begin{document}

   \thesaurus{6(02.01.2; 08.09.2; 08.13.1; 08.14.1; 08.18.1; 13.25.5)
             } 
   \title{The nature of RX~J0052.1-7319}

   \offprints{Present address: Astronomisches Institut der 
   Universit\"at Bonn, Auf dem H\"ugel 71, D-53121 Bonn, Germany}

   \author{P. Kahabka\inst{1}}

   \institute{$^1$~Astronomical Institute, University of Amsterdam, 
   Kruislaan 403, NL-1098~SJ Amsterdam, The Netherlands}

   \date{Received 22 November 1999 / accepted 30 December 1999}
 
   \maketitle
   \markboth{P. Kahabka: The nature RX~J0052.1-7319}{}
 
   \begin{abstract}
The nature of the X-ray source RX~J0052.1-7319 is discussed from 
observational data obtained from {\sl ROSAT} observations performed in
1995 and 1996. An accurate position is derived from {\sl ROSAT HRI}
observations of the source performed in 1995. The 6\asec error circle
contains two {\sl OGLE} microlensing optical variables of which one has
previously been identified with a 14.5 mag Be-type star in the Small 
Magellanic Cloud. During the October 1996 observation RX~J0052.1-7319 was 
found to be extremely bright (with a count rate of $\rm \sim1.1\pm0.1\ 
s^{-1}$) and 15.3$\pm$0.1~second X-ray pulsations have been discovered 
during this observation. This would indicate for a high-mass X-ray binary 
nature of the source. During the 1995 observation the X-ray source detected
at the position of RX~J0052.1-7319 was a factor $\sim$200 fainter. 
The corresponding luminosity has changed from $\sim5.2\times10^{37}\ erg\ 
s^{-1}$ to $\sim2.6\times10^{35}\ erg\ s^{-1}$ assuming SMC membership of
the source. It is unclear whether the so-far unidentified second optical 
variable contributes to the X-ray flux of the source.

      \keywords{Accretion -- stars: individual: RX~J0052.1-7319 -- 
                stars: magnetic fields -- stars: neutron -- stars: 
                rotation -- X-rays: stars
               }
   \end{abstract}

%

\section{Introduction}

RX~J0052.1-7319 (= 1E~0050.3-7335) has been discovered during {\sl Einstein}
observations (e.g. Seward and Mitchell 1981). But the nature of the source 
has not been determined making use of these {\sl Einstein} observations. 
The source has been found to coincide with the nebular complex DEM~70 in the 
SMC (Davies, Elliot \& Meaburn 1976). RX~J0052.1-7319 has been found as 
spectrally hard and highly variable X-ray source in the {\sl ROSAT PSPC} X-ray
survey of Kahabka \& Pietsch (1996) and it is Number 84 in the {\sl ROSAT PSPC}
X-ray catalog of Kahabka et al. (1999).

Due to its spectral hardness and the observed time variability of the X-ray 
flux it has been classified as a persistent and highly variable X-ray source
and a candidate X-ray binary system in the Small Magellanic Cloud by Kahabka
\& Pietsch (1996).
 
The detection of X-ray pulsations for this source have not been reported for
a long time and the nature of the source remained unclear. But recently 
15.3 sec X-ray pulsations have been discovered during {\sl ROSAT HRI} and 
{\sl BATSE} observations performed in Nov/Dec 1996 (Lamb et al. 1999). This
fact indicated for an X-ray pulsar associated with this source.

The source is contained in the catalog of X-ray sources detected with the
{\sl ASCA} satellite in the field of the SMC (Yokogawa et al. 1999). It has 
been found to be a comparatively weak X-ray source and no pulsations have 
been found for this source from the {\sl ASCA} observations. 
  
Searches for an optical counterpart in a 10\asec error circle for the X-ray 
source have been performed by Israel \& Stella (1999). They found a B-type 
star with R=14.54$\pm$0.03 which shows H$\alpha$ emission indicating for a 
Be type nature of the star. In addition they found another object with 
R=16.05$\pm$0.05.

The R=14.5 mag star has been found to be contained in 
the {\sl OGLE} microlensing database towards the SMC by Udalski et al. 
(1999). It is a long-term variable star with quasiperiodic light variation 
of amplitude 0.13 mag in the I band. A possible period of $\sim$600-700 
days has been found for the star. But it is rather uncertain due to the 
comparable length of the used {\sl OGLE} SMC database (of 745 days, from 1997 
Jan.~17 to 1999 Feb.~1). A second variable object with V=15.9 has been found 
in the X-ray error circle which may be identical with the object found by
Israel \& Stella (1999).

Here we discuss the possible nature of RX~J0052.1-7319. We make use of two
{\sl ROSAT HRI} observations performed in 1995 and 1996 in a systematic 
program by the author to study the time variability of (candidate) X-ray 
binary systems in the Small Magellanic Cloud. First and preliminary results 
have been reported elsewhere (Kahabka 1999a,b).

%
 
\section{Observations}

The observations discussed in this paper have been performed with the 
{\sl HRI} detector of the {\sl ROSAT} satellite (Tr\"umper 1983) in 1995 and 
1996. In Table~1 a log of these observations is given. The 1995 observation 
was centered on the SMC supernova remnant SNR~0049-736 = N19 (cf. Kahabka et 
al. 1999). The 1996 observation was centered on the supersoft X-ray source 
RX~J0048.4-7332 (cf. Kahabka et al. 1994). The source was at an off-axis angle
of 5\arcmin and 21\arcmin in the 1995 and 1996 observation respectively. It 
was in the 1996 observation close to the rim of the {\sl HRI} detector and 
affected by the varying attitude of the satellite. 

    \begin{table*}
      \caption[]{Data of the {\sl ROSAT HRI} observations of RX~J0052.1-7319 
                 in 1995 and 1996 analysed in this work.}
      \begin{flushleft}
      \begin{tabular}{ccccc}
      \hline
      \hline
      \noalign{\smallskip}
 Obs. ID   &\multicolumn{2}{c}{Observation time}                                        & Exposure  & Count rate    \\
                     & [UT]                                & [JD - 2400000]             & [ksec]    & [1/s]         \\
      \noalign{\smallskip}
      \hline
      \noalign{\smallskip}
 wg500419h (and h-1) & 18-5-1995  10:23 --  5-12-1995 11:32  & 49855.93285 -- 50056.98065 & 15.6    & $\rm(5.7\pm2.5)\times10^{-3}$ \\
 wg300513h           & 19-10-1996 11:08 -- 19-10-1996 11:41  & 50375.96413 -- 50375.98681 &  1.6    & $\rm1.1\pm0.1$\\
      \noalign{\smallskip}
      \hline
      \end{tabular}
      \end{flushleft}
   \end{table*}

\section{Results}

\subsection {The X-ray position}

   \begin{figure}
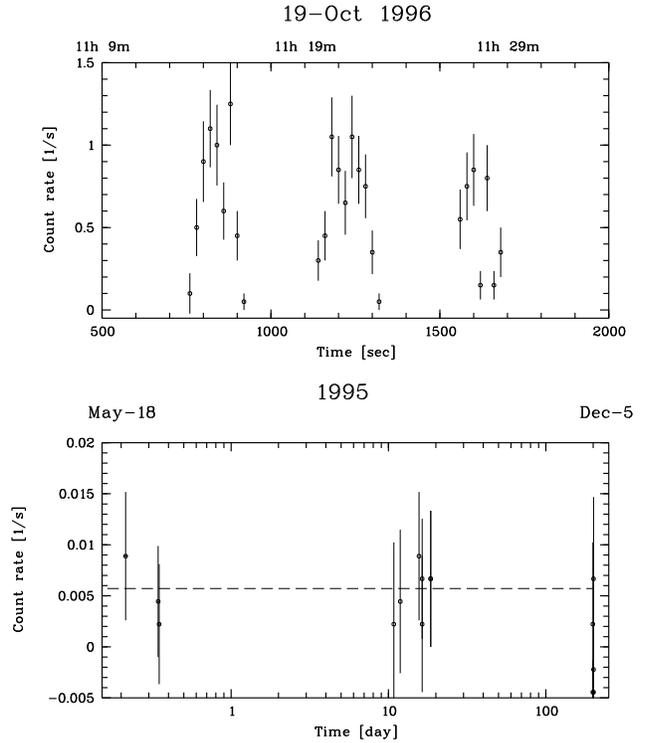

      \centering{
      \vbox{\psfig{figure=9428.f1,width=8.5cm,%
       bbllx=2.5cm,bblly=4.0cm,bburx=21.0cm,bbury=14.8cm,clip=}}\par
      \vbox{\psfig{figure=9428.f2,width=8.5cm,%
       bbllx=2.5cm,bblly=4.0cm,bburx=21.0cm,bbury=15.0cm,clip=}}\par
       }
      \caption[]{Upper panel: Light curve of RX~J0052.1-7319 for the 
                 October 1996 observation (background subtracted and 
                 corrected for off-axis vignetting). A source extraction 
                 radius of 100\asec has been used and the binsize is 
                 20~seconds. Three observation time intervals of 
                 $\sim$200~second each can clearly be seen. The large
                 variation in count rate is due to the fact that the
                 source is seen at the very edge of the field of view of 
                 the HRI. Due to the wobble of the detector and the
                 variable aspect of the satellite the source was 
                 temporarily to a large degree outside of detector. 
                 Therefore the mean count rate was low.
                 Lower panel: Light curve of RX~J0052.1-7319 for the 
                 May--Dec 1995 observation (background subtracted and 
                 corrected for vignetting). A source extraction radius of 
                 50\asec has been used and the binsize is 7.5~minutes.} 
         \label{FigLicu}
    \end{figure}

During the 1995 {\sl HRI} observation the source had a {\sl HRI} count rate
of $\sim5.6\pm2.5\times10^{-3}\ s^{-1}$ and was quite faint. It was in the 
central field of the detector and an accurate position could be determined. 
The observation extended over more than half a year (cf. Fig.~1). Three
time intervals have been analyzed independently to constrain the X-ray 
position and to determine the positional uncertainty due to a variable 
satellite aspect. As there are 
two time variable optical counterparts in the previously reported {\sl ROSAT 
PSPC} 11'' error box of RX~J0052.1-7319 (Kahabka \& Pietsch 1996) a more 
accurate {\sl HRI} position may allow to determine the association of the 
X-ray source to either of these sources. As the nature of the optically 
fainter source seems not to be constrained (e.g. in terms of a stellar source 
or a background AGN) it is also not clear whether it is a detectable X-ray 
source. In principle both objects may contribute to the observed X-ray source.

    \begin{table}
      \caption[]{X-ray position of RX~J0052.1-7319 determined from 
                 {\sl ROSAT HRI} observations in 1995 and 1996. The 90\% 
                 confidence error radius is $\sim$6\asec for the 1995 
                 observations and in the range $\sim$(5-30)\asec for the 
                 1996 observation respectively.}
      \begin{flushleft}
      \begin{tabular}{ccc}
      \hline
      \hline
      \noalign{\smallskip}
 Time interval [1995]              & R.A. (2000) & Decl. (2000) \\
      \noalign{\smallskip}
      \hline
      \noalign{\smallskip}
     \multicolumn{3}{c}{1995 observation}                       \\
      \noalign{\smallskip}
     18-May 10:19 -- 18-May 18:56  & 0:52:14.9   & -73:19:15    \\
     29-May 06:03 -- 30-May 08:43  & 0:52:16.1   & -73:19:14    \\
      3-Jun 01:36 --  5-Jun 23:53  & 0:52:15.8   & -73:19:11    \\
      \noalign{\smallskip}
      \hline
      \noalign{\smallskip}
     \multicolumn{3}{c}{1996 observation}                       \\
      \noalign{\smallskip}
     19-Oct 11:08 -- 19-Oct 11:41  & 0:52:14.1   & -73:19:16    \\
      \noalign{\smallskip}
      \hline
      \end{tabular}
      \end{flushleft}
   \end{table}

From the May--June 1995 {\sl ROSAT} {\sl HRI} observation a mean position 
R.A. = 0$^{\rm h}$52$^{\rm m}$15$^{\rm s}$.5, Decl. = -73$^{\rm o}$19'14" 
(equinox 2000.0; $\pm$4\asec at 90\% confidence) has been reported by Kahabka 
(1999a).

We derive from the three observations (cf. Tab.~2) an average position 
R.A. = 0$^{\rm h}$52$^{\rm m}$15.$^{\rm s}$.6, Decl. = -73$^{\rm o}$19'13" 
(equinox 2000.0) which is in agreement with the position derived by Kahabka 
(1999a). From the mean deviation we derive an error radius of $\sim$4\asec 
(at 90\% confidence). Assuming that this is the positional error due to the 
uncertain satellite attitude we constrain the total (statistical and 
systematic) positional error at 90\% confidence to 6\asec. 

The source is observed during the 1996 observation at a large off-axis angle 
of 21\amin. The 50\% power radius of the {\sl HRI} point-spread function is
considerable (32\asec) at such an off-axis angle but the position may be 
more accurately determined due to the central core of the point-spread 
function. The statistical 90\% error radius determined with a maximum
likelihood analysis in {\sl EXSAS} is $\sim$5\asec. The positional uncertainty
taking systematic errors into account may be between these two limits. Still
the position of the source found in the 1996 observation agrees with the
position of the source found in the 1995 observation. For comparison the 
position derived for the symbiotic nova RX~J0048.4-7332 during the same 
observation deviates $\sim$5\asec from the optical position (Morgan, 1992).

\subsection{X-ray pulsations}

We have searched for the 15.3 sec pulsations reported in this source by Lamb
et al. (1999) in the 1996 data. The event times have been projected 
from the spacecraft to the solar-system barycenter with standard {\it EXSAS} 
software (Zimmermann et al. 1994). In addition the photon event table has
been screened by selecting the wobble phase interval (0.8,1.2) using
standard {\sl EXSAS} software. This procedure has been applied as the 
point-spread function of the source is heavily affected by the very outer 
edge of the detector. Due to the satellite wobble the point-spread function 
of the source was temporarily outside of the detector. By applying this 
selection we screened the data and accepted only time intervals when the 
point-spread function was to a large fraction within the detector.
The source count rate is strongly affected by this effect. From Fig~1 we see 
that the maximum effective background subtracted count rate was large, i.e. 
$\sim1.1\pm0.1\ s^{-1}$. Assuming that this is the count rate when the source
was to a large degree inside the detector we conclude that the source has 
about this count rate. Interestingly this is about 1.5$\times$ the count rate 
as reported from the {\sl ROSAT HRI} observations of Lamb et al. (1999) which 
were performed $\sim$1-2 months after our observation. Apparently the peak of 
the outburst occurred either during our observation or even earlier. 
Assuming a spectrum with an absorbing column density of $\rm 3\ 10^{21}\ 
H-atoms\ cm^{-2}$, a photon index of 1.0, a {\sl HRI} count rate of
$\rm 1.1\ s^{-1}$ corresponds to a flux of $\rm 9.6\ 10^{-11}\ erg\ cm^{-2}\
s^{-1}$ and to an absorbed luminosity of $\rm 4.1\ 10^{37}\ erg\ s^{-1}$
and to an unabsorbed luminosity of $\rm 5.2\ 10^{37}\ erg\ s^{-1}$.

A best period of 15.3$\pm$0.1 sec has been derived from the 1996 data 
(cf. Fig~.2). The period is not very accurately constrained as the observation
was short (1.5~ksec) and additional screening of the data reduced the 
effective exposure time to 0.47~ksec. We used 6 and 4 phase bins respectively 
and the pulse profile is given in Fig~2. The period uncertainty as determined 
from the relation $\rm \delta P = P/(T_{\rm obs}\times N_{\rm bin})$ (with the 
effective exposure time $T_{\rm obs}$ and the number of phase bins 
$\rm N_{\rm bin}$) would be $\rm \sim\ 10^{-2}$~seconds. The significance of 
the period detection is $\rm \sim6\sigma$. The 1996 October~19 observation was
performed about one month before the outburst reported by Lamb et al. (1999) 
from the {\sl ROSAT} {\sl HRI} and {\sl BATSE} observations.

   \begin{figure}
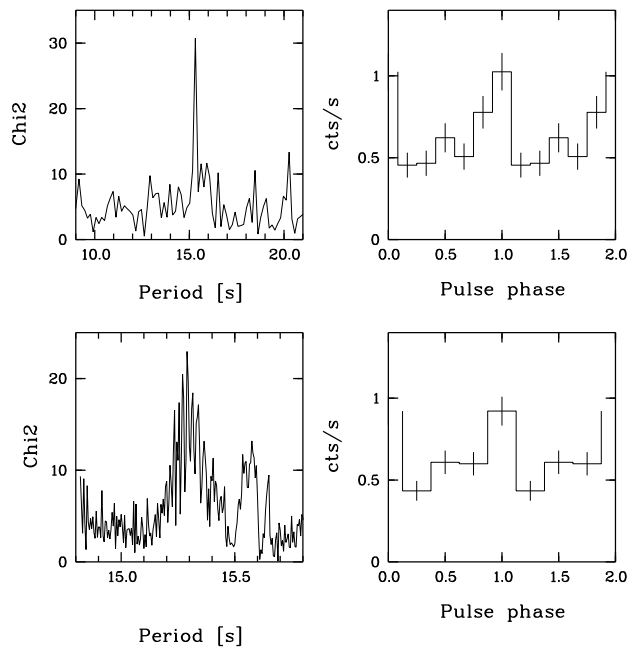

      \centering{
      \vbox{\psfig{figure=9428.f3,width=8.5cm,%
       bbllx=2.0cm,bblly=1.7cm,bburx=18.5cm,bbury=10.0cm,clip=}}\par
      \vbox{\psfig{figure=9428.f4,width=8.5cm,%
       bbllx=2.0cm,bblly=1.7cm,bburx=18.5cm,bbury=10.0cm,clip=}}\par
       }
      \caption[]{$\chi^2$ distribution for the 15.3 sec pulse period search 
                 applied to period data of RX~J0052.1-7319 in October 1996.
                 Upper panel for 6 phase bin and wider period range and lower
                 panel for 4 phase bins and more narrow period range.}
         \label{FigChi}
    \end{figure}

    \begin{table}
      \caption[]{Pulse period search in the 1996 October 19 {\sl ROSAT HRI} 
                 observation of RX~J0052.1-7319. The effective exposure 
                 time and count rate as used in the time window for the
                 period search are given. Ony time intervals when the 
                 point-spread function was within the detector due to
                 the wobble of the attitude have been chosen.}
      \begin{flushleft}
      \begin{tabular}{lc}
      \hline
      \hline
      \noalign{\smallskip}
 Observation                & 300513h           \\
 date                       & 1996 Oct 19       \\
 Effective expos. (ksec)    & 0.47              \\
 Pulse period (s)           & 15.3(1)           \\
 $\chi^2$ / dof             & 31/5              \\
 Effective count rate [1/s] & 0.65              \\
      \noalign{\smallskip}
      \hline
      \end{tabular}
      \end{flushleft}
   \end{table}

\section{Discussion}

Kahabka \& Pietsch (1996) have argued that RX~J0052.1-7319 may be a persistent
and highly variable X-ray source as it has already been observed during \ein 
observations (Seward \& Mitchell 1981; Inoue et al. 1983; Wang \& Wu 1992). 
The source has been detected in \ros {\sl PSPC} observations performed in 
October 1991 and April 1992, and with the {\sl HRI} performed in May to 
December 1995 and October 1996. The count rate was varying by a large factor 
$>$100. The source has not always been detected during {\sl ROSAT} 
observations. Cowley \& Schmidtke (1997) report that the source was not 
detected during {\sl HRI} observations performed in April, May, and October 
1994. They derive an upper limit to the count rate of $< 0.004\ s^{-1}$ which 
is still in the range of count rates derived for the 1995 observation 
($\rm 0.0057\pm0.0025\ s^{-1}$). Analysis of archival {\sl PSPC} observations 
performed in December 1992 and May 1993 do not reveal a significant detection 
of the source which has been at large off-axis angles of 44\amin and 35\amin 
respectively. The $\rm2\sigma$ upper limit {\sl PSPC} 
count rate is $\sim2\times10^{-3}\ s^{-1}$ for the May 1993 observation and a 
factor of 10 lower than the {\sl PSPC} count rate derived for the April 1992 
observation (Kahabka \& Pietsch 1996). The rise in X-ray flux 
(of $\sim0.01\ counts\ s^{-1}\ d^{-1}$) during the April 1992 observation
could be due to the onset of an X-ray outburst which might have happened 
around August 1993 and decayed till November 1993. This would give an  outburst
repetition time of 3.2~years (or 1200~days). This behaviour is in agreement 
with a High-Mass Be-type X-ray binary system which undergoes high states 
during periastron passage of the neutron star. 

The source was indeed observed during the 1996 observations for about 50~days 
(from 1996 October 19 till December 9) in an outburst with an {\sl HRI}
count rate of $\rm 0.7-1.2\ s^{-1}$. If this outburst is due to the periastron
passage of the neutron star than constraints can be derived for the orbital
period which has to be considerably longer.

RX~J0052.1-7319 is in the OGLE SMC6 field and an about 2 year observational
database exists for this source. Two OGLE detected variable stars are within
the reported X-ray error circle of RX~J0052.1-7319, SMC~SC6~99923 and
SMC~SC6~99991 (Udalski 1999). The first is a long-term variable star with 
quasiperiodic light variation of amplitude 0.13 in the I band (the mean I 
magnitude is 14.5). The possible period is 600-700~days, but it is uncertain 
due to the length of the database. The second optical variable has an I 
magnitude in the range 16.1 to 15.7 with a pronounced linear rise in the I 
magnitude during a 200 day period. It may be identical with the object 
reported by Israel \& Stella 1999 (IAU Circ. No. 7101), which remains so-far
unidentified. No $\rm H\alpha$ line emission has been detected from this
object. It could in principle be a background AGN. This may explain the
fact that no X-ray pulsations have been detected with {\sl ASCA} although an
X-ray source has been detected at the position of RX~J0052.1-7319.
Both objects are within the 6\asec {\sl HRI} error circle.

It appears to be likely that RX~J0052.1-7319 is associated with the 14.5 mag
Be-type star and is a Be-type transient. But it cannot be excluded that the
X-ray source is confused by a second near-by source, e.g. a time variable 
background stellar source or AGN. If it is one X-ray source then it also has 
to be understood why it is active in X-rays for nearly two months. If the 
source follows the relation between rotation period and orbital period found 
by Corbet for the galactic Be-type transients (Corbet 1986) then the 
15.3~second rotation period would correspond to an orbital period of 
$\sim$40~days. If the duration of the 1996 outburst is indeed $\sim$50~days 
then this would argue against activity related to periastron passage for such 
a short orbital period. 

Another possibility is that the high-mass star associated with RX~J0052.1-7319
periodically undergoes ``eruptions'' or outbursts (of duration a few months) 
during which it efficiently transfers mass towards the neutron star companion
(cf. Marlborough 1997). If the star settles back to its normal configuration 
the mass-transfer reduces. Such a scenario would still be consistent with an 
orbital period of $\sim$40~days in this system as the duration of the X-ray
outburst would be determined by the duration of the outburst of the star. 
A Be star undergoing periodic outbursts in our Galaxy is $\lambda$~Eri. In a 
recent work Mennickent et al. (1998) found for this system an outburst 
repetition period of 469~days (or 939~days). The duration of the outburst 
is $\sim$120~days. 

The similarity of the long-period lightcurve of $\lambda$~Eri and of the 
{\sl OGLE} microlensing light curve of the 15.9 star SMC~SC6~99991 in the 
error box of RX~J0052.1-7319 is striking. As it is not clear whether this 
object is a stellar source (in the SMC) or a time variable background AGN (cf. 
Kawaguchi et al. 1998) it could be related to RX~J0052.1-7319. Note that the
fact that no $\rm H\alpha$ emission is observed (Israel \& Stella 1999) also
refers to $\lambda$~Eri (Mennickent et al. 1998). One problem with this 
identification may be that the observed 1996 X-ray outburst occured outside
the ``optical outburst'' although ``projecting'' the SMC~SC6~99991 light
curve onto the $\lambda$~Eri light curve a repetition period of 
$\sim$3.5~years would be obtained which is quite close to the X-ray outburst
period of RX~J0052.1-7319 estimated from the X-ray observations.

\section{Summary}
For RX~J0052.1-7319 15.3 sec pulsations have been detected in {\sl ROSAT HRI}
X-ray observations performed in October 1996. The count rate of 
$\rm 1.1\pm0.1\ s^{-1}$ observed during this observation is the highest 
reported so far for this source. The corresponding X-ray luminosity is 
$\sim5\times 10^{37}\ erg\ s^{-1}$ for SMC distance. The position of the 
X-ray source coincides with a 14.5~mag Be-type star in the SMC and a 
fainter $\sim$15.9~mag object which remains so-far unidentified. Both objects 
have been found to be variable with a timescale of a few hundred days and 
are contained in the {\sl OGLE} microlensing database. It is unclear which 
is the optical counterpart of RX~J0052.1-7319 or if even both objects 
contribute to the observed X-ray flux.

\acknowledgements
The {\sl ROSAT} project is supported by the Max-Planck-Gesellschaft and the 
Bundesministerium f\"ur Forschung und Technologie (BMFT). This research was 
supported in part by the Netherlands Organisation for Scientific Research 
(NWO) through Spinoza Grant 08-0 to E.P.J. van den Heuvel. I thank Lex Kaper
for discussions and W. Kundt for reading the manuscript. I thank the referee
R.C. Lamb for useful comments.




\begin{thebibliography}{}
\bibitem{} Corbet R.H.D., 1986, in: The Evolution of Galactic X-Ray Binaries,
           eds. J. Tr\"umper et al., Reidel Pub. Co., p. 63
\bibitem{} Cowley A.P., Schmidtke P.C., McGrath T.K., et al., 1997, PASP 109,
           21
\bibitem{} Davies R.D., Elliot K.H., Meaburn J., 1976, Mem. R. Astr. Soc. 81, 
           89
\bibitem{} Inoue H., Koyama K., Tanaka Y., 1983, in: IAU Symposium 101,
           Supernova Remnants and Their X-Ray Emission, eds. Danziger J.,
           Gorenstein P., Dordrecht: Reidel, p. 535
\bibitem{} Israel G.L., Stella L., 1999, IAU Circ. No. 7101
\bibitem{} Kahabka P., Pietsch W., Hasinger G., 1994, A\&A 288, 538
\bibitem{} Kahabka P., Pietsch W., 1996, A\&A 312, 919
\bibitem{} Kahabka P., 1999a, IAU Circ. No. 7082
\bibitem{} Kahabka P., 1999b, IAU Circ. No. 7087
\bibitem{} Kahabka P., Pietsch W., Filipovic M.D., Haberl F., 1999, A\&AS 
           136, 81
\bibitem{} Kawaguchi T., Mineshige S., 1998, ApJ 504, 671
\bibitem{} Lamb R.C., Prince T.A., Macomb D.J., et al., 1999, 
           IAU Circ. No. 7081
\bibitem{} Marlborough J.M., 1997, A\&A 317, L17
\bibitem{} Mennickent R.E., Sterken C., Vogt N., 1998, A\&A 330, 631
\bibitem{} Morgan D.H., 1992, MNRAS 258, 639
\bibitem{} Seward F.D. \& Mitchell M. 1981, ApJ 243, 736
\bibitem{} Tr\"umper J. 1983, Adv. Spa. Res. 2, 241
\bibitem{} Udalski A., 1999, IAU Circ. No. 7105
\bibitem{} Wang Q., Wu X., 1992, ApJS 78, 391
\bibitem{} Yokogawa J., Imanishi K., Tsujimoto M., et al., 1999, ApJ 
           (subm.)
\bibitem{} Zimmermann H.U. et al. 1994, MPE report 257
\vskip 1.0 cm
\vfill\eject
\end{thebibliography}
\end{document}